\documentclass[nofootinbib,10pt,twocolumn,pra,showpacs,tightenlines, showkeys, english,aps]{revtex4}
\usepackage[T1]{fontenc}
\usepackage[latin9]{inputenc}
\usepackage{color}
\usepackage{array}
\usepackage{amsmath}
\usepackage{graphicx}

\usepackage{epsfig,graphicx,times}
\usepackage{amstext}
\usepackage{amsmath}            
\usepackage{amssymb}            
\usepackage{graphicx}           
\usepackage{latexsym}
\usepackage{bm}
\usepackage{color}
\usepackage{amssymb}
\usepackage{epstopdf}
\usepackage{epsfig}
\usepackage{caption}
\makeatletter

\usepackage{epsfig}\usepackage{bm}\usepackage{epsfig}\@ifundefined{definecolor}
 {\usepackage{color}}{}

\def\be{\begin{equation}}
\def\ee{\end{equation}}
\def\bea{\begin{eqnarray}}
\def\eea{\end{eqnarray}}

\@ifundefined{showcaptionsetup}{}{%
 \PassOptionsToPackage{caption=false}{subfig}}
\usepackage{subfig}
\makeatother

\usepackage{babel}

\begin{document}

\title{{\normalsize Higher order nonclassicalities in a codirectional nonlinear
optical coupler: Quantum entanglement, squeezing and antibunching}}

\author{Kishore Thapliyal$^{a}$, Anirban Pathak$^{a,b,}$\footnote{Email: anirban.pathak@gmail.com,  Phone: +91 9717066494}, Biswajit Sen$^{c}$, and Jan ${\rm \check{Perina}}$$^{b,d}$ 
}
\affiliation{
$^{a}$Jaypee Institute of Information Technology, A-10, Sector-62,
Noida, UP-201307, India\\
$^{b}$RCPTM, Joint Laboratory of Optics of Palacky University and
Institute of Physics of Academy of Science of the Czech Republic,
Faculty of Science, Palacky University, 17. listopadu 12, 771 46 Olomouc,
Czech Republic\\
$^{c}$Department of Physics, Vidyasagar Teachers' Training College,
Midnapore-721101, India\\
$^{d}$Department of Optics, Palacky University, 17. listopadu 12,
771 46 Olomouc, Czech Republic}

\begin{abstract}

{\normalsize Higher order nonclassical properties of fields propagating
through a codirectional asymmetric nonlinear optical coupler which
is prepared by combining a linear wave guide and a nonlinear (quadratic)
wave guide operated by second harmonic generation are studied. A completely
quantum mechanical description is used here to describe the system.
Closed form analytic solutions of Heisenberg's equations of motion
for various modes are used to show the existence of higher order antibunching,
higher order squeezing, higher order two-mode and multi-mode entanglement
in the asymmetric nonlinear optical coupler. It is also shown that
nonclassical properties of light can transfer from a nonlinear wave
guide to a linear wave guide.}{\normalsize \par}

\end{abstract}



\maketitle

\section{{\normalsize Introduction}}

Several new applications of nonclassical states have been reported
in recent past \cite{CV-qkd-hillery,teleportation of coherent state,antibunching-sps,Ekert protocol,Bennet1993,densecoding}.
For example, applications of squeezed state are reported in implementation
of continuous variable quantum cryptography \cite{CV-qkd-hillery},
teleportation of coherent states \cite{teleportation of coherent state},
etc., antibunching is shown to be useful in building single photon
sources \cite{antibunching-sps}, entangled state has appeared as
one of the main resources of quantum information processing as it
is shown to be essential for the implementation of a set of protocols
of discrete \cite{Ekert protocol} and continuous variable quantum
cryptography \cite{CV-qkd-hillery}, quantum teleportation \cite{Bennet1993},
dense-coding \cite{densecoding}, etc. As a consequence of these recently
reported applications, generation of nonclassical states in various
quantum systems emerged as one of the most important areas of interest
in quantum information theory and quantum optics. Several systems
are already investigated and have been shown to produce entanglement
and other nonclassical states (see \cite{pathak-perina,pathak-PRA}
and references therein). However, experimentally realizable simple
systems that can be used to generate and manipulate nonclassical states
are still of much interest. One such experimentally realizable and
relatively simple system is nonlinear optical coupler. Nonlinear optical
couplers are of specific interest because they can be easily realized
using optical fibers or photonic crystals and the amount of nonclassicality
present in the output field can be controlled by controlling the interaction
length and the coupling constant. Further, recently Matthews et al.
have experimentally demonstrated manipulation of multi-photon entanglement
in quantum circuits constructed using waveguides \cite{Nature09}.
Quantum circuits implemented by them can also be viewed as optical
coupler based quantum circuits as in their circuits waveguides are
essentially combined to form couplers. In another interesting application,
Mandal and Midda have shown that universal irreversible gate library
(NAND gate) can be built using nonlinear optical couplers \cite{Mandal-Midda}.
Mandal and Midda's work essentially showed that in principle a classical
computer can be built using optical couplers. These facts motivated
us to systematically investigate the possibility of observation of
nonclassicality in nonlinear optical couplers. Among different possible
nonlinear optical couplers one of the simplest systems is a codirectional
asymmetric nonlinear optical coupler that is prepared by combining
a linear wave guide and a nonlinear (quadratic) wave guide operated
by second harmonic generation. Waveguides interact with each other
through evanescent wave and we may say that transfer of nonclassical
effect from the nonlinear wave guide to the linear one happens through
evanescent wave. Present paper aims to study various higher order
nonclassical properties of this specific optical coupler with specific
attention to entanglement.

It is interesting to note that several nonclassical properties of
optical couplers are studied in past (see \cite{perina-review} for
a review). For example, photon statistics, phase properties and squeezing
in codirectional and contradirectional Kerr nonlinear coupler are
studied with fixed and varying linear coupling constant \cite{Kerr-1,kerr-2,Kerr-3,kerr-4,kerr-5},
photon statistics of Raman and Brillouin coupler \cite{Raman and Brilloin}
and parametric coupler \cite{parametric} is studied in detail, photon
statistics and other nonclassical properties of asymmetric \cite{assymetric-strong-pump,assymetric-2,Mandal-Perina,co-and-contra,contra-1,nonclassical-input1}
and symmetric \cite{nonclassical-input1,nonclasssical input2} directional
nonlinear coupler is investigated for various conditions such as strong
pump \cite{assymetric-strong-pump}, weak pump \cite{Mandal-Perina},
phase mismatching \cite{co and contra with phase mismatch} for codirectional
\cite{Mandal-Perina,co-and-contra,co and contra with phase mismatch}
and contradirectional \cite{co-and-contra,contra-1,co and contra with phase mismatch}
propagation of classical (coherent) and nonclassical \cite{nonclassical-input1,nonclasssical input2}
input modes. However, almost all the earlier studies were limited
to the investigation of lower order nonclassical effects (e.g., squeezing
and antibunching) either under the conventional short-length approximation
\cite{contra-1} or under the parametric approximation where a pump
mode is assumed to be strong and treated classically as a c-number
\cite{co and contra with phase mismatch}. Only a few discrete efforts
have recently been made to study higher order nonclassical effects
and entanglement in optical couplers \cite{leonoski1,thermal ent-kerr,kerr-lionoski-miranowicz,GBS,higher-order},
but even these efforts are limited to Kerr nonlinear coupler. For
example, in 2004, Leonski and Miranowicz reported entanglement in
Kerr nonlinear coupler \cite{kerr-lionoski-miranowicz}, subsequently
entanglement sudden death \cite{leonoski1} and thermally induced
entanglement \cite{thermal ent-kerr} are reported in Kerr nonlinear
coupler. Amplitude squared (higher order) squeezing is also reported
in Kerr nonlinear coupler \cite{higher-order}. However, neither any
effort has yet been made to rigorously study the higher order nonclassical
effects in nonlinear optical coupler in general nor a serious effort
has been made to study entanglement in nonlinear optical couplers
other than Kerr nonlinear coupler.  Keeping these facts in mind in
the present paper we aim to study higher order nonclassical effects
(e.g., higher order antibunching, squeezing and entanglement) in codirectional
nonlinear optical coupler.

Remaining part of the paper is organized as follows. In Section \ref{sec:The-model-and}
we briefly describe the Hamiltonian that describes the model of the
asymmetric nonlinear optical coupler studied here and perturbative
solutions of equations of motion corresponding to different field
modes present in the Hamiltonian. In Section \ref{sec:Criteria-of-nonclassicality}
we list a set of criteria of nonclassicality with special attention
to those kind of nonclassicalities that are never explored for asymmetric
nonlinear optical coupler. In Section \ref{sec:Nonclassicality-in-codirectional}
we use the criteria described in the previous section to illustrate
the nonclassical characters of various field modes present in the
asymmetric nonlinear optical coupler. Specifically, we have reported
higher order squeezing, antibunching, and entanglement. Finally, the
paper is concluded in Section \ref{sec:Conclusions}.

\section{{\normalsize The model and the solutions\label{sec:The-model-and}}}

\begin{figure}
\begin{centering}
\includegraphics[scale=0.8]{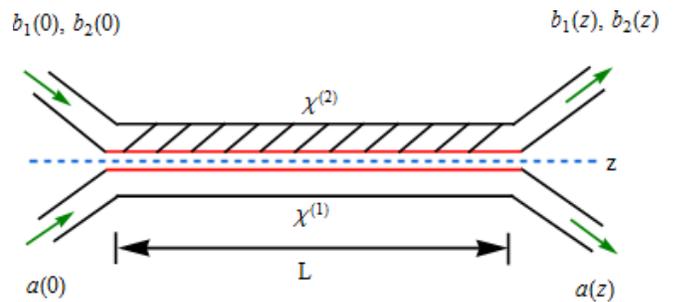}
\par\end{centering}

\caption{\label{fig:Schematic-diagram}Schematic diagram of a codirectional
asymmetric nonlinear optical coupler prepared by combining a linear
wave guide with a nonlinear (quadratic) wave guide operated by second
harmonic generation. The fields involved are described by the corresponding
annihilation operators, as shown; $L$ is the interaction length.}
\end{figure}

An asymmetric nonlinear optical coupler is schematically shown in
Fig. \ref{fig:Schematic-diagram}. We are interested in nonclassical
properties of this coupler. From Fig. \ref{fig:Schematic-diagram}
we can clearly see that a linear wave guide is combined with a nonlinear
one with $\chi^{(2)}$ nonlinearity to from the asymmetric coupler.
As the $\chi^{(2)}$ medium can produce second harmonic generation,
we may say that the coupler is operated by second harmonic generation.
The linear waveguide carries the electromagnetic field characterized
by the bosonic field annihilation (creation) operator $a\,(a^{\dagger})$.
On the other hand, the field operators $b_{i}\,(b_{i}^{\dagger})$
correspond to the nonlinear medium. Further, $b_{1}(k_{1})$ and $b_{2}(k_{2})$
denote annihilation operators (wave vectors) for fundamental and second
harmonic modes, respectively. Now the nonlinear momentum operator
in the interaction picture for this coupler can be written as \begin{equation}
G=-\hbar kab_{1}^{\dagger}-\hbar\Gamma b_{1}^{2}b_{2}^{\dagger}\exp(i\Delta kz)\,+{\rm h.c}.\,,\label{eq:1}\end{equation}
where ${\rm h.c.}$ stands for the Hermitian conjugate and $\Delta k=|2k_{1}-k_{2}|,$
denotes the phase mismatch between the fundamental and second harmonic
beams. The parameters $k$ and $\Gamma$ are the linear and nonlinear
coupling constants and are proportional to the linear $(\chi^{(1)})$
and nonlinear $(\chi^{(2)})$ susceptibilities, respectively. The
value of $\chi^{(2)}$ is considerably smaller than $\chi^{(1)}$
(typically $\chi^{(2)}/\chi^{(1)}\,\simeq10^{-6})$ and as a consequence
$\Gamma\ll k$ unless an extremely strong pump is present. The model
is elaborately discussed by some of the present authors in their earlier
publications \cite{perina-review,assymetric-2,Mandal-Perina}. Specifically,
in Ref. \cite{assymetric-2} single mode and intermodal squeezing,
antibunching and subshot noise was studied using analytic expressions
of spatial evolution of field operators obtained by short-length solution
of the Heisenberg's equations of motion corresponding to (\ref{eq:1}).
The validity of short length solution used in Ref. \cite{assymetric-2}
was strictly restricted by the condition $\Gamma z\ll1.$ Later on
Sen and Mandal developed a perturbative solution technique \cite{bsen1}
that can solve Heisenberg's equations of motion for $\Gamma z\ll1.$
Sen-Mandal technique was subsequently used in Ref. \cite{Mandal-Perina}
to obtain spatial evolution of field operators corresponding to (\ref{eq:1})
and to study single mode and intermodal squeezing and antibunching.
Interestingly, in \cite{Mandal-Perina} some nonclassical characters
of asymmetric nonlinear optical coupler were observed which were not
observed in the earlier investigations \cite{perina-review,assymetric-2}
performed using short-length solution. This was indicative of the
fact that the Sen-Mandal perturbative method provides better solution%
\footnote{In fact, short-length (time) solution can be obtained as a special
case of Sen-Mandal perturbative solution. %
} for the study of nonclassical properties. The same fact is observed
in other optical systems, too (\cite{pathak-PRA} and references therein).
However, neither entanglement nor any of the higher order nonclassical
properties were studied in earlier papers. Keeping these facts in
mind we have used the solution reported in Ref. \cite{Mandal-Perina}
to study the higher order nonclassicalities. 

In Ref. \cite{Mandal-Perina} closed form analytic expressions for
evolution of field operators valid up to linear power of the coupling
coefficient $\Gamma$ were obtained as follows: \begin{equation}
\begin{array}{lcl}
a(z) & = & f_{1}a(0)+f_{2}b_{1}(0)+f_{3}b_{1}^{\dagger}(0)b_{2}(0)+f_{4}a^{\dagger}(0)b_{2}(0),\\
b_{1}(z) & = & g_{1}a(0)+g_{2}b_{1}(0)+g_{3}b_{1}^{\dagger}(0)b_{2}(0)+g_{4}a^{\dagger}(0)b_{2}(0),\\
b_{2}(z) & = & h_{1}b_{2}(0)+h_{2}b_{1}^{2}(0)+h_{3}b_{1}(0)a(0)+h_{4}a^{2}(0),\end{array}\label{eq:2}\end{equation}
where \begin{equation}
\begin{array}{lcl}
f_{1} & = & g_{2}=\cos|k|z,\\
f_{2} & = & -g_{1}*=-\frac{ik^{*}}{|k|}\sin|k|z,\\
f_{3} & = & \frac{2k^{*}\Gamma^{*}}{4|k|^{2}-(\Delta k)^{2}}\left[G_{-}f_{1}+\frac{f_{2}}{k^{*}}\left\{ \Delta k-\frac{2|k|^{2}}{\Delta k}G_{-}\right\} \right],\\
f_{4} & = & \frac{4k^{*2}\Gamma^{*}}{\Delta k\left[4|k|^{2}-(\Delta k)^{2}\right]}G_{-}f_{1}+\frac{2k^{*}\Gamma^{*}}{\left[4|k|^{2}-(\Delta k)^{2}\right]}G_{+}f_{2},\\
g_{3} & = & \frac{2\Gamma^{*}k}{\left[4|k|^{2}-(\Delta k)^{2}\right]}G_{+}f_{2}-\frac{2\Gamma^{*}\left(2|k|^{2}-(\Delta k)^{2}\right)f_{1}}{\Delta k\left[4|k|^{2}-(\Delta k)^{2}\right]}G_{-},\\
g_{4} & = & \frac{4\Gamma^{*}|k|^{2}}{\Delta k\left[4|k|^{2}-(\Delta k)^{2}\right]}f_{2}-\frac{2\Gamma^{*}\left(2|k|^{2}-(\Delta k)^{2}\right)}{\Delta k\left[4|k|^{2}-(\Delta k)^{2}\right]}\\
 & \times & \left(G_{+}-1\right)f_{2}+\frac{2k^{*}\Gamma^{*}}{\left[4|k|^{2}-(\Delta k)^{2}\right]}G_{-}f_{1},\\
h_{1} & = & 1,\\
h_{2} & = & \frac{\Gamma G_{-}^{*}}{2\Delta k}-\frac{i\Gamma}{2\left[4|k|^{2}-(\Delta k)^{2}\right]}\left[2|k|\left(G_{+}^{*}-1\right)\sin2|k|z\right.\\
 & - & \left.i\Delta k\left(1-\left(G_{+}^{*}-1\right)\cos2|k|z\right)\right],\\
h_{3} & = & \frac{-\Gamma|k|}{k^{*}\left[4|k|^{2}-(\Delta k)^{2}\right]}\left[i\Delta k\left(G_{+}^{*}-1\right)\sin2|k|z\right.\\
 & + & \left.2|k|\left(1-\left(G_{+}^{*}-1\right)\cos2|k|z\right)\right],\\
h_{4} & =- & \frac{\Gamma|k|^{2}G_{-}^{*}}{2k^{*^{2}}\Delta k}-\frac{i\Gamma|k|^{2}}{2k^{*^{2}}\left[4|k|^{2}-(\Delta k)^{2}\right]}\left[2|k|\left(G_{+}^{*}-1\right)\right.\\
 & \times & \left.\sin2|k|z-i\Delta k\left(1-\left(G_{+}^{*}-1\right)\cos2|k|z\right)\right],\end{array}\label{eq:3}\end{equation}
where $G_{\pm}=\left(1\pm\exp(-i\Delta kz)\right).$ In what follows
we will use these closed form analytic expressions of the field operators
to investigate the spatial evolution of entanglement and some higher
order nonclassical characteristics of the field modes. We will not
discuss the usual nonclassical characters such as squeezing and antibunching
as they are already discussed in Ref. \cite{Mandal-Perina}.

\section{{\normalsize Criteria of nonclassicality \label{sec:Criteria-of-nonclassicality}}}

A state having negative or highly singular (more singular than $\delta$-function)
Glauber-Sudarshan $P$-function is referred to as a nonclassical state
as it cannot be expressed as a classical mixture of coherent states.
$P$-function provides us an essential as well as sufficient criterion
for detection of nonclassicality. However, $P$-function is not directly
experimentally measurable. Consequently, several operational criteria
for nonclassicality are proposed in last 50 years. A large number
of these criteria are expressed as inequalities involving expectation
values of functions of annihilation and creation operators. This implies
that Eqns. (\ref{eq:2})-(\ref{eq:3}) provide us the sufficient mathematical
framework required to study the nonclassical properties of the codirectional
asymmetric nonlinear optical coupler. As mentioned above we are interested
in higher order nonclassical properties of radiation field. In quantum
optics and quantum information higher order nonclassical properties
of bosons (e.g., higher order Hong-Mandel squeezing, higher order
antibunching, higher order sub-Poissonian statistics, higher order
entanglement, etc.) are often studied (\cite{generalized-higher order}
and references therein). Until recent past studies on higher order
nonclassicalities were predominantly restricted to theoretical investigations.
However, a bunch of exciting experimental demonstrations of higher
order nonclassicalities are recently reported \cite{Maria-PRA-1,Maria-2,higher-order-PRL}.
Specifically, existence of higher order nonclassicality in bipartite
multi-mode states produced in a twin-beam experiment is recently demonstrated
by Allevi, Olivares and Bondani \cite{Maria-PRA-1} using a new criterion
for higher order nonclassicality introduced by them. They also showed
that detection of weak nonclassicalities is easier with their higher
order criterion of nonclassicality as compared to the existing lower
order criteria \cite{Maria-PRA-1}. This observation was consistent
with the earlier theoretical observation of Pathak and Garcia \cite{HOAwithMartin}
that established that depth of nonclassicality in higher order antibunching
increases with the order. The possibility that higher order nonclassicality
may be more useful in identifying the weak nonclassicalities have
considerably increased the interest of the quantum optics community
on the higher order nonclassical characters of bosonic fields. In
the remaining part of this section we list a set of criteria of higher
order nonclassicalities and in the following section we study the
possibility of satisfying those criteria in the codirectional asymmetric
nonlinear optical coupler.

\subsection{{\normalsize Higher order squeezing}}

Higher order squeezing is usually studied using two different approaches
\cite{HIlery-amp-sq,Hong-Mandel1,HOng-mandel2}. In the first approach
introduced by Hillery in 1987 \cite{HIlery-amp-sq} reduction of variance
of an amplitude powered quadrature variable for a quantum state with
respect to its coherent state counter part reflects nonclassicality.
In contrast in the second type of higher order squeezing introduced
by Hong and Mandel in 1985 \cite{Hong-Mandel1,HOng-mandel2}, higher
order squeezing is reflected through the reduction of higher order
moments of usual quadrature operators with respect to their coherent
state counterparts. In the present paper we have studied higher order
squeezing using Hillery's criterion of amplitude powered squeezing.
Specifically, Hillery introduced amplitude powered quadrature variables
as \begin{equation}
Y_{1,a}=\frac{a^{k}+\left(a^{\dagger}\right)^{k}}{2}\label{eq:quadrature-power1}\end{equation}
 and \begin{equation}
Y_{2,a}=i\left(\frac{\left(a^{\dagger}\right)^{k}-a^{k}}{2}\right).\label{eq:quadrature-power2}\end{equation}
As $Y_{1,a}$ and $Y_{2,a}$ do not commute we can obtain uncertainty
relation and a condition of squeezing. For example, for $k=2$, Hillery's
criterion for amplitude squared squeezing is described as \begin{equation}
A_{i,a}=\left\langle \left(\Delta Y_{i,a}\right)^{2}\right\rangle -\left\langle N_{a}+\frac{1}{2}\right\rangle <0,\label{eq:criterion-amplitude squared}\end{equation}
where $i\in\{1,2\}.$

\subsection{{\normalsize Higher order antibunching}}

Since 1977 signatures of higher order nonclassical photon statistics
in different optical systems of interest have been investigated by
some of the present authors using criterion based on higher order
moments of number operators (cf. Ref. \cite{perina-review} and Chapter
10 of \cite{Perina-Book} and references therein). However, higher
order antibunching (HOA) was not specifically discussed, but it was
demonstrated there for degenerate and nondegenerate parametric processes
in single and compound signal-idler modes, respectively and for Raman
scattering in compound Stokes-anti-Stokes mode up to $n=5$. Further,
it was shown that the HOA is deeper with increasing $n$ occurring
on a shorter time interval in parametric processes whereas different
order HOA occurs on the same time interval in Raman scattering. A
specific criterion for HOA was first introduced by C. T. Lee \cite{C T Lee}
in 1990 using higher order moments of number operator. Initially,
HOA was considered to be a phenomenon that appears rarely in optical
systems, but in 2006, some of the present authors established that
it is not really a rare phenomenon \cite{HOAis not rare}. Since then
HOA is reported in several quantum optical systems (\cite{generalized-higher order}
and references therein) and atomic systems \cite{with bishu Arxive}.
However, no effort has yet been made to study HOA in optical couplers.
Thus the present study of HOA in asymmetric nonlinear optical coupler
is first of its kind and is expected to lead to similar observations
in other type of optical couplers. Before we proceed further, we would
like to note that signature of HOA can be observed through a bunch
of equivalent but different criteria, all of which can be interpreted
as modified Lee criterion. In what follows we will use following simple
criterion of $(n-1)^{th}$ order single mode antibunching introduced
by Pathak and Garcia \cite{HOAwithMartin} \begin{equation}
\begin{array}{lcl}
D_{a}(n-1)=\left\langle a^{\dagger n}a^{n}\right\rangle -\left\langle a^{\dagger}a\right\rangle ^{n} & < & 0.\end{array}\label{hoa}\end{equation}
Here $n=2$ corresponds to the usual antibunching and $n\geq3$ refers
to the higher order antibunching.

\subsection{{\normalsize Entanglement and higher order entanglement}}

There exist several inseparability criteria (\cite{NJP-GS-Ashoka}
and references therein) that are expressed in terms of expectation
values of field operators and thus suitable for study of entanglement
dynamics within the frame-work of the present approach. Among these
criteria Duan et al.'s criterion \cite{duan} which is usually referred
to as Duan's criterion, Hillery-Zubairy criterion I and II (HZ-I and
HZ-II) \cite{HZ-PRL,HZ2007,HZ2010} have received more attention because
of various reasons, such as computational simplicity, experimental
realizability and their recent success in detecting entanglement in
various optical, atomic and optomechanical systems (\cite{pathak-PRA,with bishu Arxive}
and references therein). To begin with we may note that the first
inseparability criterion of Hillery and Zubairy, i.e., HZ-1 criterion
of inseparability is described as \begin{equation}
\begin{array}{lcl}
\left\langle N_{a}N_{b}\right\rangle  & -\left|\left\langle ab^{\dagger}\right\rangle \right|^{2}< & 0,\end{array}\label{hz1}\end{equation}
whereas the second criterion of Hillery and Zubairy, i.e., HZ-II criterion
is given by \begin{equation}
\begin{array}{lcl}
\left\langle N_{a}\right\rangle \left\langle N_{b}\right\rangle  & -\left|\left\langle ab\right\rangle \right|^{2}< & 0.\end{array}\label{hz2}\end{equation}
 The other criterion of inseparability to be used in the present paper
is Duan et al.'s criterion which is described as follows \cite{duan}:
\begin{equation}
\begin{array}{lcl}
d_{ab}=\left\langle \left(\Delta u_{ab}\right)^{2}\right\rangle +\left\langle \left(\Delta v_{ab}\right)^{2}\right\rangle -2 & < & 0,\end{array}\label{duan}\end{equation}
 where \begin{equation}
\begin{array}{lcl}
u_{ab} & = & \frac{1}{\sqrt{2}}\left\{ \left(a+a^{\dagger}\right)+\left(b+b^{\dagger}\right)\right\}, \\
v_{ab} & = & -\frac{i}{\sqrt{2}}\left\{ \left(a-a^{\dagger}\right)+\left(b-b^{\dagger}\right)\right\} .\end{array}\label{eq:duan-2}\end{equation}
Clearly our analytic solution (\ref{eq:2})-(\ref{eq:3}) enables
us to investigate intermodal entanglement in asymmetric nonlinear
optical coupler using all the three inseparability criteria described
above. As all these three inseparability criteria are only sufficient
(not necessary), a particular criterion may fail to identify entanglement
detected by another criterion. Keeping this fact in mind, we use all
these criteria to study the intermodal entanglement in asymmetric
nonlinear optical coupler. The criteria described above can only detect
bi-partite entanglement of lowest order. As possibility of generation
of entanglement in asymmetric nonlinear optical coupler is not discussed
earlier we have studied the spatial evolution of intermodal entanglement
using these lower order inseparability criteria. However, to be consistent
with the focus of the present paper, we need to investigate the possibility
of observing higher order entanglement, too. For that purpose we require
another set of criteria for detection of higher order entanglement.
All criteria for detection of multi-partite entanglement are essentially
higher order criteria \cite{higher order multiparty1,higher order multyparty2,higher order multyparty3}
as they reveal some higher order correlation. Interestingly, there
exist higher order inseparability criteria for detection of higher
order entanglement in bipartite case, too. Specifically, Hillery-Zubairy
introduced two criteria for intermodal higher order entanglement \cite{HZ-PRL}
as follows \begin{equation}
E_{ab}^{m,n}=\left\langle \left(a^{\dagger}\right)^{m}a^{m}\left(b^{\dagger}\right)^{n}b^{n}\right\rangle -\left\vert \left\langle a^{m}\left(b^{\dagger}\right)^{n}\right\rangle \right\vert ^{2}<0,\label{hoe-criteria}\end{equation}
and\begin{equation}
E_{ab}^{'m,n}=\left\langle \left(a^{\dagger}\right)^{m}a^{m}\rangle\langle\left(b^{\dagger}\right)^{n}b^{n}\right\rangle -\left\vert \left\langle a^{m}b^{n}\right\rangle \right\vert ^{2}<0.\label{hoe-criteria-1}\end{equation}
Here $m$ and $n$ are non-zero positive integers and lowest possible
values of $m$ and $n$ are $m=n=1$ which reduces (\ref{hoe-criteria})
and (\ref{hoe-criteria-1}) to usual HZ-I criterion (i.e., (\ref{hz1}))
and HZ-II criterion (i.e., (\ref{hz2})), respectively. Thus these
two criteria are generalized version of well known lower order criteria
of Hillery and Zubairy and we may refer to (\ref{hoe-criteria}) and
(\ref{hoe-criteria-1}) as HZ-I criterion and HZ-II criterion respectively
in analogy to the lowest order cases. A quantum state will be referred
to as (bipartite) higher order entangled state if it is found to satisfy
(\ref{hoe-criteria}) and/or (\ref{hoe-criteria-1}) for any choice
of integer $m$ and $n$ satisfying $m+n\geq3.$ The other type of
higher order entanglement i.e., multi-partite entanglement can be
detected in various ways. In the present paper we have used a set
of multi-mode inseparability criteria introduced by Li et al. \cite{Ent condition-multimode}.
Specifically, Li et al. have shown that a three-mode quantum state
is not bi-separable in the form $ab_{1}|b_{2}$ (i.e., compound mode
$ab_{1}$ is entangled with the mode $b_{2}$) if the following inequality
holds for the three-mode system 

\begin{equation}
E_{ab_{1}|b_{2}}^{m,n,l}=\langle\left(a^{\dagger}\right)^{m}a^{m}\left(b_{1}^{\dagger}\right)^{n}b_{1}^{n}\left(b_{2}^{\dagger}\right)^{l}b_{2}^{l}\rangle-|\langle a^{m}b_{1}^{n}(b_{2}^{\dagger})^{l}\rangle|^{2}<0,\label{eq:tripartite ent1}\end{equation}
where $m,\, n,\, l$ are positive integers and annihilation operators
$a,b_{1},b_{2}$ correspond to the three modes. A quantum state satisfying
the above inequality is referred to as $ab_{1}|b_{2}$ entangled state.
Three mode inseparability criterion can be written in various alternative
forms. For example, an alternative criterion for detection of $ab_{1}|b_{2}$
entangled state is \cite{Ent condition-multimode} \begin{equation}
E_{ab_{1}|b_{2}}^{^{\prime}m,n,l}=\langle\left(a^{\dagger}\right)^{m}a^{m}\left(b_{1}^{\dagger}\right)^{n}b_{1}^{n}\rangle\langle\left(b_{2}^{\dagger}\right)^{l}b_{2}^{l}\rangle-|\langle a^{m}b_{1}^{n}b_{2}^{l}\rangle|^{2}<0.\label{eq:tripartite ent2}\end{equation}
Similarly, one can define criteria for detection of $a|b_{1}b_{2}$
and $b_{1}|ab_{2}$ entangled states and use them to obtain criterion
for detection of fully entangled tripartite state. For example, using
(\ref{eq:tripartite ent1}) and (\ref{eq:tripartite ent2}) respectively
we can write that the three modes of our interest are not bi-separable
in any form if any one of the following two sets of inequalities are
satisfied simultaneously \begin{equation}
E_{ab_{1}|b_{2}}^{1,1,1}<0,\, E_{a|b_{1}b_{2}}^{1,1,1}<0,\, E_{b_{1}|b_{2}a}^{1,1,1}<0,\label{eq:fully enta 0}\end{equation}

\begin{equation}
E_{ab_{1}|b_{2}}^{^{\prime}1,1,1}<0,\, E_{a|b_{1}b_{2}}^{^{\prime}1,1,1}<0,\, E_{b_{1}|b_{2}a}^{^{\prime}1,1,1}<0.\label{eq:fully enta 1}\end{equation}
 Further, for a fully separable pure state we always have \begin{equation}
|\langle ab_{1}b_{2}\rangle|=|\langle a\rangle\langle b_{1}\rangle\langle b_{2}\rangle|\leq\left[\langle N_{a}\rangle\langle N_{b_{1}}\rangle\langle N_{b_{2}}\rangle\right]^{\frac{1}{2}}.\label{eq:fully ent2}\end{equation}
Thus a 3-mode pure state that violates (\ref{eq:fully ent2}) (i.e.,
satisfies $\langle N_{a}\rangle\langle N_{b_{1}}\rangle\langle N_{b_{2}}\rangle-|\langle ab_{1}b_{2}\rangle|^{2}<0)$
and simultaneously satisfies either (\ref{eq:fully enta 0}) or (\ref{eq:fully enta 1})
is a fully entangled state as it is neither fully separable nor bi-separable
in any form.

\section{{\normalsize Nonclassicality in codirectional optical coupler \label{sec:Nonclassicality-in-codirectional}}}

Using the perturbative solutions (\ref{eq:2})-(\ref{eq:3}) we can
obtain time evolution of various operators that are relevant for the
detection of nonclassical characters. For example, we may use (\ref{eq:2})-(\ref{eq:3})
to obtain the number operators for various field modes as follows
\begin{widetext}
\begin{equation}
\begin{array}{lcl}
N_{a} & = & a^{\dagger}a=|f_{1}|^{2}a^{\dagger}(0)a(0)+|f_{2}|^{2}b_{1}^{\dagger}(0)b_{1}(0)+\left[f_{1}^{*}f_{2}a^{\dagger}(0)b_{1}(0)+f_{1}^{*}f_{3}a^{\dagger}(0)b_{1}^{\dagger}(0)b_{2}(0)\right.\\
 & + & \left.f_{1}^{*}f_{4}a^{\dagger2}(0)b_{2}(0)+f_{2}^{*}f_{3}b_{1}^{\dagger2}(0)b_{2}(0)+f_{2}^{*}f_{4}b_{1}^{\dagger}(0)a^{\dagger}(0)b_{2}(0)+{\rm h.c.}\right],\end{array}\label{eq:na}\end{equation}
 \begin{equation}
\begin{array}{lcl}
N_{b_{1}} & = & b_{1}^{\dagger}b_{1}=|g_{1}|^{2}a^{\dagger}(0)a(0)+|g_{2}|^{2}b_{1}^{\dagger}(0)b_{1}(0)+\left[g_{1}^{*}g_{2}a^{\dagger}(0)b_{1}(0)+g_{1}^{*}g_{3}a^{\dagger}(0)b_{1}^{\dagger}(0)b_{2}(0)\right.\\
 & + & \left.g_{1}^{*}g_{4}a^{\dagger2}(0)b_{2}(0)+g_{2}^{*}g_{3}b_{1}^{\dagger2}(0)b_{2}(0)+g_{2}^{*}g_{4}b_{1}^{\dagger}(0)a^{\dagger}(0)b_{2}(0)+{\rm h.c.}\right],\end{array}\label{eq:nb}\end{equation}
 \begin{equation}
N_{b_{2}}=b_{2}^{\dagger}b_{2}=b_{2}^{\dagger}(0)b_{2}(0)+\left[h_{2}b_{2}^{\dagger}(0)b_{1}^{2}(0)+h_{3}b_{2}^{\dagger}(0)b_{1}(0)a(0)+h_{4}b_{2}^{\dagger}(0)a^{2}(0)+{\rm h.c.}\right].\label{eq:nb2}\end{equation}
\end{widetext}
The average value of the number of photons in the modes $a,$ $b_{1}$
and $b_{2}$ may now be calculated with respect to a given initial
state. We assume that initial state is a product of three coherent
states: $|\alpha\rangle|\beta\rangle|\gamma\rangle,$ where $|\alpha\rangle,\,|\beta\rangle$
and $|\gamma\rangle$ are eigen kets of annihilation operators $a,\, b_{1}$and
$b_{2}$, respectively. Field operator $a(0)$ operating on such a
multi-mode coherent state yields a complex eigenvalue $\alpha.$ Specifically,
\begin{equation}
a(0)|\alpha\rangle|\beta\rangle|\gamma\rangle=\alpha|\alpha\rangle|\beta\rangle|\gamma\rangle,\label{2bcd}\end{equation}
where $|\alpha|^{2},\,|\beta|^{2},\,|\gamma|^{2}$ is the number of
input photons in the field mode $a,\, b_{1}$ and $b_{2}$, respectively.
For a spontaneous process, the complex amplitudes should satisfy $\beta=\gamma=0$
and $\alpha\ne0.$ Whereas, for a stimulated process, the complex
amplitudes are not necessarily zero and it would be physically reasonable
to choose $\alpha>\beta>\gamma.$ In what follows, in all the figures
(except Fig. \ref{fig:HOA}) that illustrate the existence of nonclassical
character in asymmetric nonlinear optical coupler we have chosen $\alpha=5,\,\beta=2,\,\gamma=1.$

\subsection{{\normalsize Higher order squeezing}}

\begin{figure}
\begin{centering}
\includegraphics[scale=0.5]{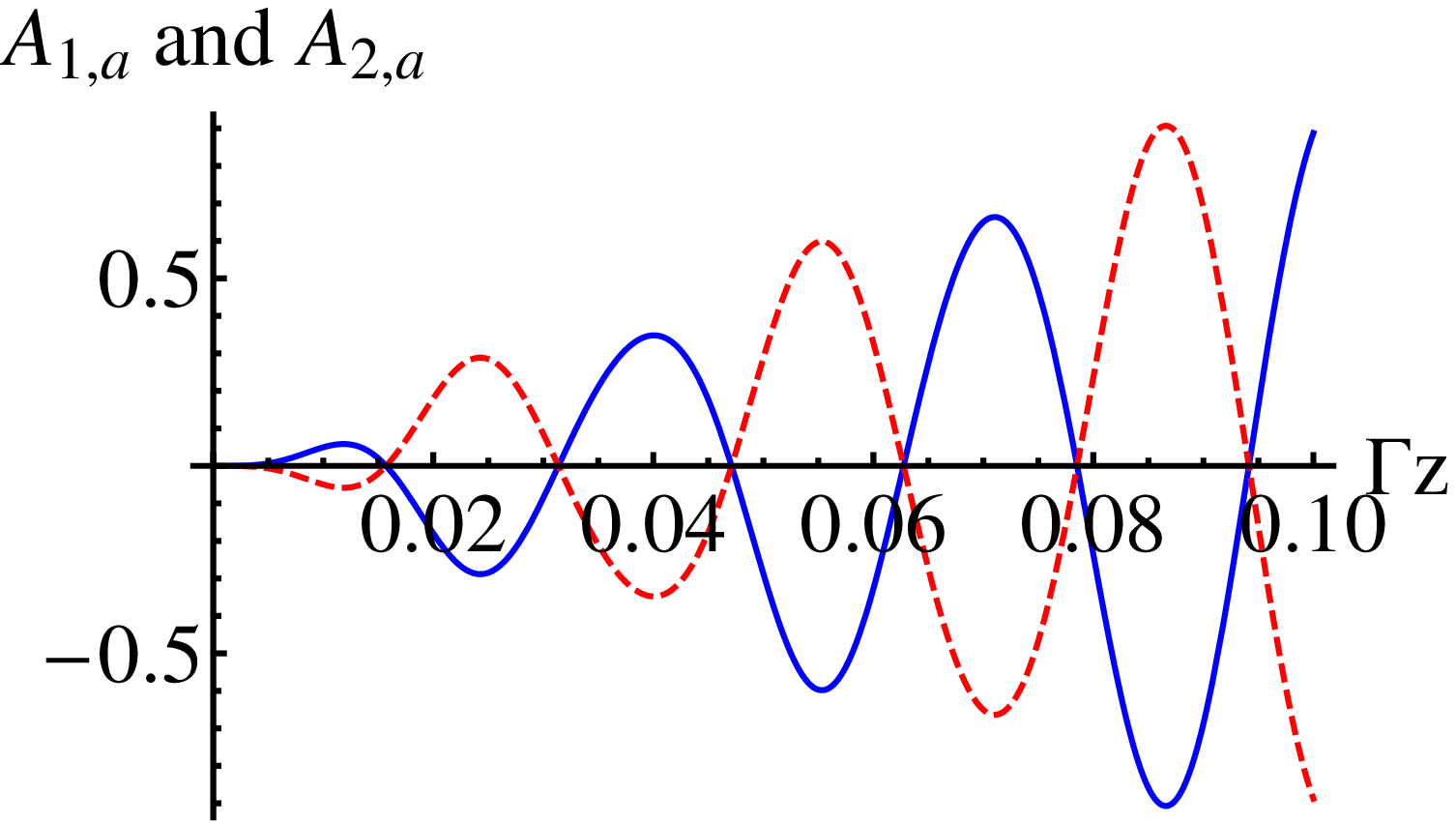}
\par\end{centering}

\centering{}\includegraphics[scale=0.5]{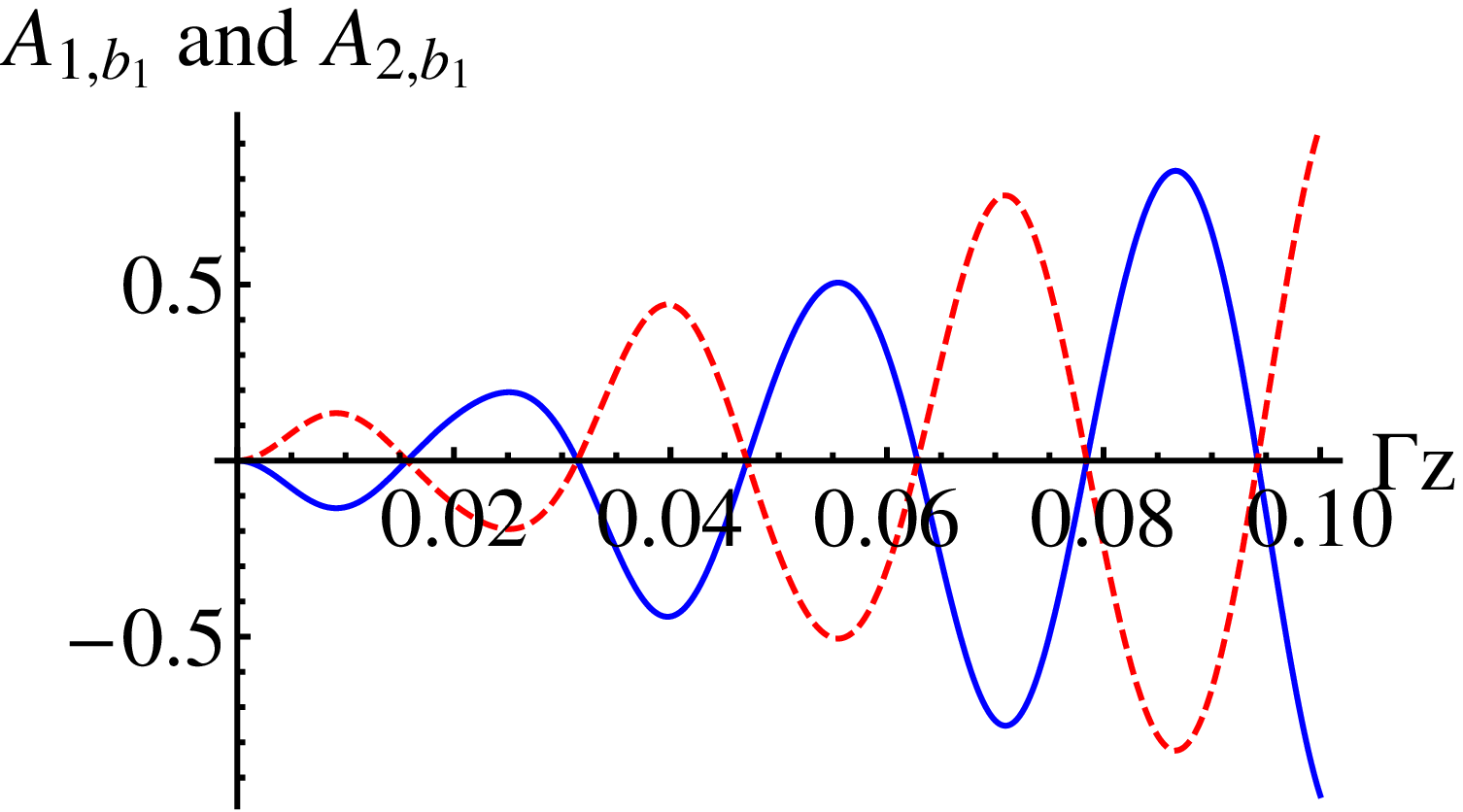}\caption{{\large \label{fig:Amplitude-squared-squeezing}}(color online) Amplitude
squared squeezing is observed in modes $a$ (top) and $b_{1}$ (bottom)
for the initial state $|\alpha\rangle|\beta\rangle|\gamma\rangle$
with $k=0.1,\,\Gamma=0.001,\,\Delta k=10^{-4},\,\alpha=5,\beta=2,\gamma=1.$
Negative parts of the solid line represents squeezing in quadrature
variable $Y_{1,a}$ ($Y_{1,b_{1}})$ and that of the dashed line represents
squeezing in quadrature variable $Y_{2,a}$ ($Y_{2,b_{1}})$. }
\end{figure}

Using Eqs. (\ref{eq:2})-(\ref{eq:3}), (\ref{eq:na})-(\ref{eq:nb2})
in the criterion of amplitude squared squeezing (\ref{eq:criterion-amplitude squared}),
we obtain \begin{equation}
\begin{array}{lcl}
\left[\begin{array}{c}
A_{1,a}\\
A_{2,a}\end{array}\right] & = & \pm\left[\left(f_{1}f_{4}+f_{2}f_{3}\right)\left(f_{1}^{2}\alpha^{2}\gamma+f_{2}^{2}\beta^{2}\gamma\right.\right.\\
 & + & \left.\left.f_{1}f_{2}\alpha\beta\gamma\right)+{\rm c.c.}\right],\end{array}\label{eq:asq-1}\end{equation}

\begin{equation}
\begin{array}{lcl}
\left[\begin{array}{c}
A_{1,b_{1}}\\
A_{2,b_{1}}\end{array}\right] & = & \pm\left[\left(g_{1}g_{4}+g_{2}g_{3}\right)\left(g_{1}^{2}\alpha^{2}\gamma+g_{2}^{2}\beta^{2}\gamma\right.\right.\\
 & + & \left.\left.g_{1}g_{2}\alpha\beta\gamma\right)+{\rm c.c.}\right],\end{array}\label{eq:as-q-2}\end{equation}
 and \begin{equation}
\begin{array}{lcl}
\left[\begin{array}{c}
A_{1,b_{2}}\\
A_{2,b_{2}}\end{array}\right] & = & 0.\end{array}\label{eq:as-q-3}\end{equation}
Clearly we don't obtain any signature of amplitude squared squeezing
in $b_{2}$ mode using the present solution and mode $a$ $(b_{1})$
should always show amplitude squared squeezing in one of the quadrature
variables as both $A_{1,a}$ and $A_{2,a}$ ($A_{1,b_{1}}$ and $A_{2,b_{2}}$)
cannot be positive simultaneously. To investigate the possibility
of amplitude squared squeezing in further detail in modes $a$ and
$b_{1}$ we have plotted the spatial variation of $A_{i,a}$ and $A_{i,b_{1}}$
in Fig. \ref{fig:Amplitude-squared-squeezing}. Negative regions of
these two plots clearly illustrate the existence of amplitude squared
squeezing in both $a$ and $b_{1}$ modes. 

\subsection{{\normalsize Higher order antibunching}}

We have already described the condition of HOA as (\ref{hoa}). Now
using Eqns. (\ref{eq:2})-(\ref{eq:3}), (\ref{hoa}) and (\ref{eq:na})-(\ref{eq:nb2})
we can obtain closed form analytic expressions for $D_{i}(n)$ for
various modes as follows 
\begin{widetext}
\begin{equation}
\begin{array}{lcl}
D_{a}(n) & = & ^{n}C_{2}\gamma|\left(f_{1}\alpha+f_{2}\beta\right)|^{2n-4}\left\{ \left(f_{1}\alpha+f_{2}\beta\right)^{2}\left(f_{2}^{*}f_{3}^{*}+f_{1}^{*}f_{4}^{*}\right)+{\rm c.c.}\right\} ,\end{array}\label{eq:dan}\end{equation}
\begin{equation}
\begin{array}{lcl}
D_{b_{1}}(n) & = & ^{n}C_{2}\gamma|\left(g_{1}\alpha+g_{2}\beta\right)|^{2n-4}\left\{ \left(g_{1}\alpha+g_{2}\beta\right)^{2}\left(g_{2}^{*}g_{3}^{*}+g_{1}^{*}g_{4}^{*}\right)+{\rm c.c.}\right\} ,\end{array}\label{eq:dbn}\end{equation}

\begin{equation}
D_{b_{2}}(n)=0.\label{eq:db2n}\end{equation}
\end{widetext}

\begin{figure}
\begin{centering}
\includegraphics[scale=0.4]{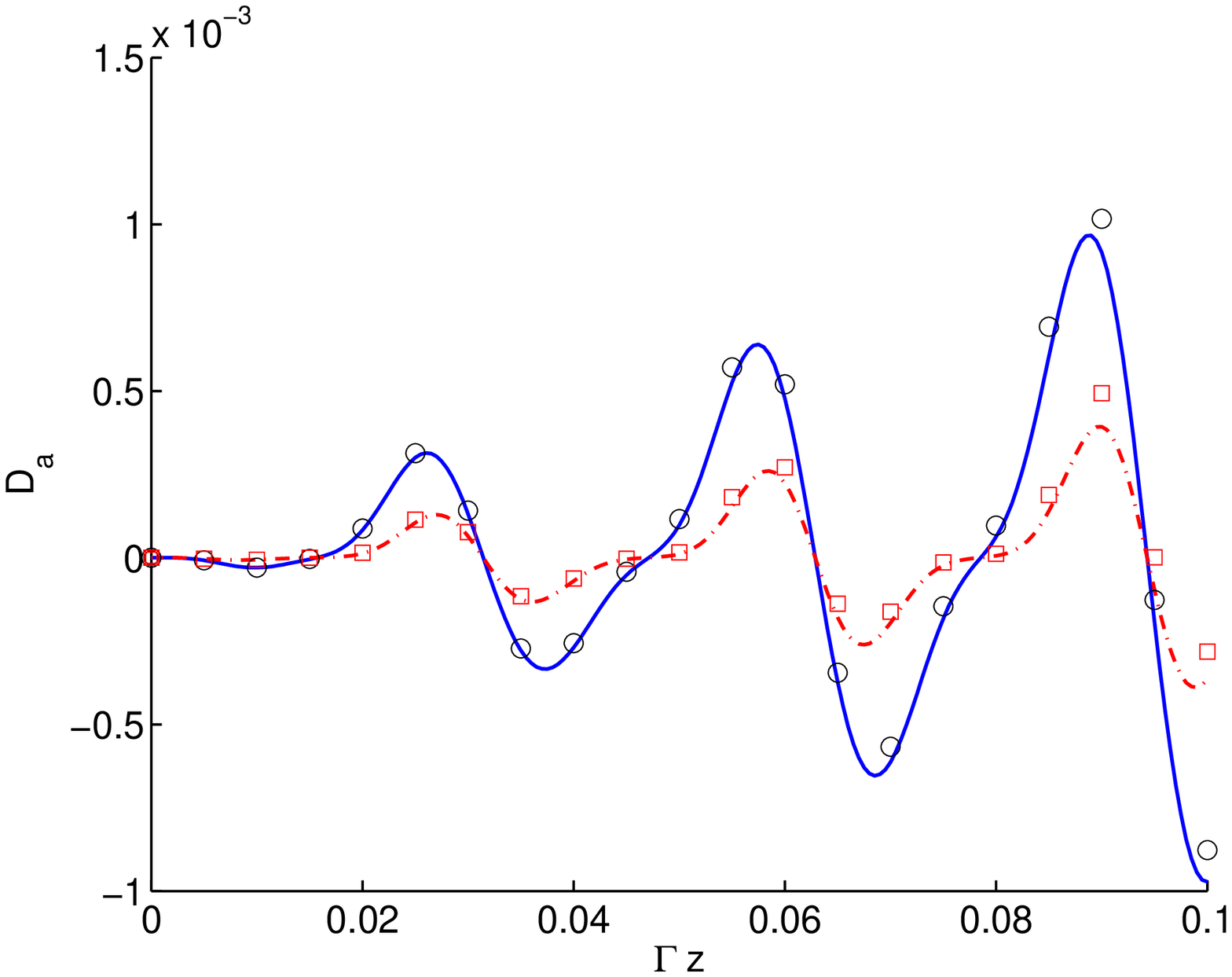}
\par\end{centering}

\centering{}\includegraphics[scale=0.4]{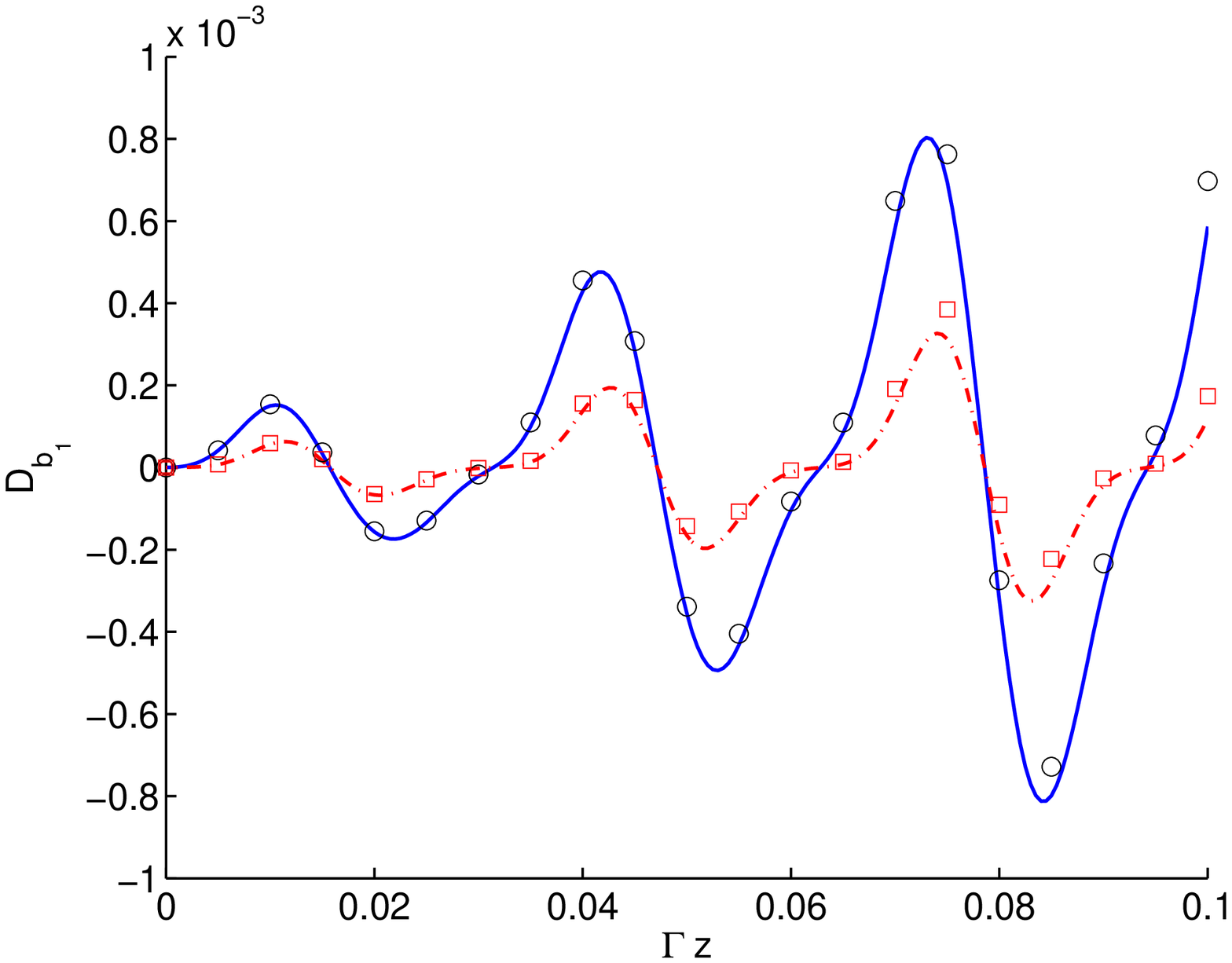}\caption{\label{fig:HOA} (color online) HOA with rescaled interaction length
$\Gamma z$ in mode $a$ (top panel) and mode $b_{1}$ (bottom
panel) for $n=3$ (smooth line) and $n=4$ (dashed line) and square
and circle are for the corresponding numerical results with the initial
state $|\alpha\rangle|\beta\rangle|\gamma\rangle$ and $k=0.1,\,\Gamma=0.001,\,\Delta k=10^{-4},\,\alpha=0.5,\,\beta=0.2,\,\gamma=0.1.$ }
\end{figure}

Clearly, the perturbative solution used here cannot detect any signature
of higher order antibunching for $b_{2}$ mode. However, in the other
two modes we observe HOA for different values of $n$ as illustrated
in Fig. \ref{fig:HOA}. In Fig. \ref{fig:HOA}, we have plotted right
hand sides of (\ref{eq:dan}) and (\ref{eq:dbn}) along with the exact
numerical results obtained by integrating the time dependent Schrodinger
equation corresponding to given Hamiltonian by using the matrix form
of the operators. Close resemblance of the exact numerical result
with the perturbative result even for higher order case clearly validates
the perturbative solution used here.

\subsection{{\normalsize Intermodal entanglement}}

To apply HZ-I criterion to investigate the existence of intermodal
entanglement between modes $a$ and $b_{1}$ i.e., compound mode $ab_{1}$
we use Eqns. (\ref{eq:2})-(\ref{eq:3}) and (\ref{eq:na})-(\ref{eq:nb2})
and obtain
\begin{equation}
\begin{array}{lcl}
E_{ab_{1}}^{1,1} & = & \langle N_{a}N_{b_{1}}\rangle-|\langle ab_{1}^{\dagger}\rangle|^{2}\\
 & = & \left(|g_{1}|^{2}f_{4}^{*}f_{1}+f_{3}^{*}f_{1}g_{2}^{*}g_{1}\right)\alpha^{2}\gamma^{*}\\
 & + & \left(|f_{1}|^{2}g_{1}^{*}g_{4}+f_{1}^{*}f_{2}g_{1}^{*}g_{3}\right)\alpha^{*2}\gamma\\
 & + & \left(|g_{2}|^{2}f_{3}^{*}f_{2}+f_{4}^{*}f_{2}g_{1}^{*}g_{2}\right)\beta^{2}\gamma^{*}\\
 & + & \left(|f_{2}|^{2}g_{2}^{*}g_{3}+f_{2}^{*}f_{1}g_{2}^{*}g_{4}\right)\beta^{*2}\gamma\\
 & + & \left(|g_{1}|^{2}-|g_{2}|^{2}\right)\left(\left(f_{4}^{*}f_{2}-f_{3}^{*}f_{1}\right)\alpha\beta\gamma^{*}\right.\\
 & - & \left.\left(g_{2}^{*}g_{4}-g_{1}^{*}g_{3}\right)\alpha^{*}\beta^{*}\gamma\right).\end{array}\label{eq:HZ-1-ab1}\end{equation}
Similarly, applying HZ-II criterion to the compound mode $ab_{1}$
we obtain

\begin{equation}
\begin{array}{lcl}
E_{ab_{1}}^{'1,1} & = & \langle N_{a}\rangle\langle N_{b_{1}}\rangle-|\langle ab_{1}\rangle|^{2}\\
 & = & -\left[\left(|g_{1}|^{2}f_{4}^{*}f_{1}+f_{3}^{*}f_{1}g_{2}^{*}g_{1}\right)\alpha^{2}\gamma^{*}\right.\\
 & + & \left.\left(|f_{1}|^{2}g_{1}^{*}g_{4}+f_{1}^{*}f_{2}g_{1}^{*}g_{3}\right)\alpha^{*2}\gamma\right.\\
 & + & \left.\left(|g_{2}|^{2}f_{3}^{*}f_{2}+f_{4}^{*}f_{2}g_{1}^{*}g_{2}\right)\beta^{2}\gamma^{*}\right.\\
 & + & \left(|f_{2}|^{2}g_{2}^{*}g_{3}+f_{2}^{*}f_{1}g_{2}^{*}g_{4}\right)\beta^{*2}\gamma\\
 & + & \left(|g_{1}|^{2}-|g_{2}|^{2}\right)\left(\left(f_{4}^{*}f_{2}-f_{3}^{*}f_{1}\right)\alpha\beta\gamma^{*}\right.\\
 & - & \left.\left.\left(g_{2}^{*}g_{4}-g_{1}^{*}g_{3}\right)\alpha^{*}\beta^{*}\gamma\right)\right].\end{array}\label{eq:hz2-ab1}\end{equation}

\begin{figure}
\begin{centering}
\includegraphics[scale=0.5]{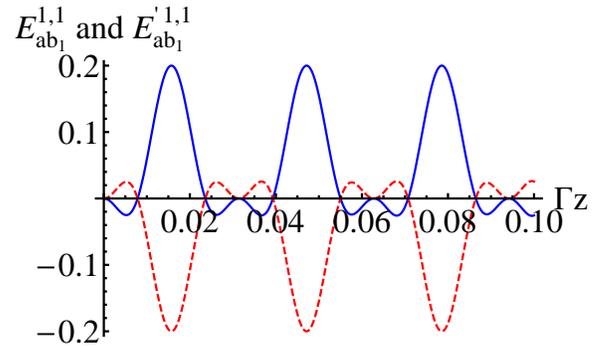}
\par\end{centering}

\caption{{\large \label{fig:ent-hz1-II}}(color online) Hillery-Zubairy criterion
I (solid line) and criterion II (dashed line) for entanglement are
showing intermodal entanglement between modes $a$ and $b_{1}.$ Here
$E_{ab_{1}}^{1,1}$ (solid line) and $E_{ab_{1}}^{\prime1,1}$ (dashed
line) are plotted with rescaled interaction length $\Gamma z$
for mode $ab_{1}$ with the initial state $|\alpha\rangle|\beta\rangle|\gamma\rangle$
and $k=0.1,\,\Gamma=0.001,\,\Delta k=10^{-4},\,\alpha=5,\beta=2,\gamma=1.$}
\end{figure}

From Eqns. (\ref{eq:HZ-1-ab1}) and (\ref{eq:hz2-ab1}) we can easily
observe that in the present case $E_{ab_{1}}^{1,1}=-E_{ab_{1}}^{'1,1}$,
which implies that at any point inside the coupler either HZ-I criterion
or HZ-II criterion would show the existence of entanglement as both
of them cannot be simultaneously positive. Thus compound mode $ab_{1}$
is always entangled inside a codirectional asymmetric optical coupler.
The same is explicitly illustrated through Fig. \ref{fig:ent-hz1-II}.
Following the same approach we investigated the existence of entanglement
in other compound modes (e.g., $ab_{2}$ and $b_{1}b_{2}),$ but both
HZ-I and HZ-II criteria failed to detect any entanglement in these
cases. However, it does not indicate that the modes are separable
as both HZ-I and HZ-II inseparability criteria are only sufficient
and not essential. Further, the perturbative analytic solution used
here is an approximate solution and in recent past we have seen several
examples where the existence of entanglement not detected by HZ criteria
is detected by Duan et al.'s criterion or vice versa \cite{pathak-PRA,with bishu Arxive}.
Keeping these facts in mind, we studied the possibilities of observing
intermodal entanglement using Duan et al.'s criterion, too, but it
failed to detect any entanglement in the present case as we obtained
\begin{equation}
d_{ab_{1}}=d_{ab_{2}}=d_{b_{1}b_{2}}=0.\label{eq:variance-duan}\end{equation}

\begin{figure}
\begin{centering}
\includegraphics[scale=0.5]{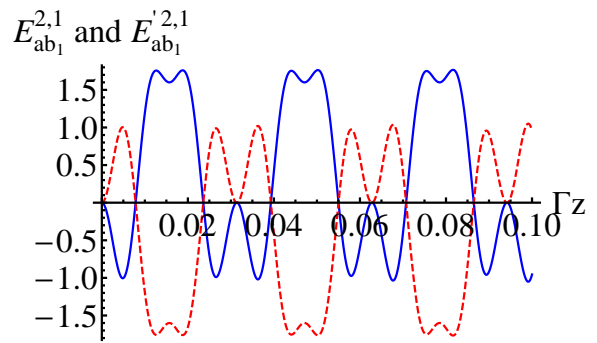}
\par\end{centering}

\caption{\textcolor{red}{\large \label{fig:hoe}}(color online) Higher order
entanglement is observed using Hillery-Zubairy criteria. Solid line
shows spatial variation of $E_{ab_{1}}^{2,1}$ and dashed line shows
spatial variation of $E_{ab_{1}}^{\prime2,1}$ with the initial state
$|\alpha\rangle|\beta\rangle|\gamma\rangle$ and $k=0.1,\,\Gamma=0.001,\,\Delta k=10^{-4},\,\alpha=5,\beta=2,\gamma=1.$ }
\end{figure}

We may now investigate the existence of higher order entanglement
using Eqns. (\ref{hoe-criteria})-(\ref{eq:fully ent2}). To begin
with we may use (\ref{eq:2})-(\ref{eq:3}) and (\ref{hoe-criteria})
to obtain \begin{equation}
\begin{array}{lcl}
E_{ab_{1}}^{m,n} & = & \langle a^{\dagger m}a^{m}b_{1}^{\dagger n}b_{1}^{n}\rangle-|\langle a^{m}b_{1}^{\dagger n}\rangle|^{2}\\
 & = & mn\left|\left(f_{1}\alpha+f_{2}\beta\right)\right|^{2m-2}\left|\left(g_{1}\alpha+g_{2}\beta\right)\right|^{2n-2}E_{ab_{1}}^{1,1}.\end{array}\label{eq:emnab1}\end{equation}
Similarly, using (\ref{eq:2})-(\ref{eq:3}) and (\ref{hoe-criteria-1})
we can obtain a closed form analytic expression for $E_{ab_{1}}^{'m,n}$
and easily observe that \begin{equation}
E_{ab_{1}}^{'m,n}=-E_{ab_{1}}^{m,n}.\label{eq:emnprimeab1}\end{equation}
Eqn. (\ref{eq:emnprimeab1}) clearly shows that higher order entanglement
between $a$ mode and $b_{1}$ mode would always be observed for any
choice of $m$ and $n$ as $E_{ab_{1}}^{m,n}$ and $E_{ab_{1}}^{'m,n}$
cannot be simultaneously positive. Using (\ref{eq:emnab1}) and (\ref{eq:emnprimeab1})
we can easily obtain analytic expressions of $E_{ab_{1}}^{2,1},$
$E_{ab_{1}}^{\prime2,1},$ $E_{ab_{1}}^{\prime1,2},$ etc. Such analytic
expressions are not reported here as existence of higher order entanglement
is clearly seen through (\ref{eq:emnprimeab1}). However, in Fig. \ref{fig:hoe} we have illustrated the spatial
evolution of $E_{ab_{1}}^{2,1}$ and $E_{ab_{1}}^{\prime2,1}$. Negative
regions of this figure clearly show the existence of higher order
intermodal entanglement in compound mode $ab_{1}.$ As expected from
(\ref{eq:emnprimeab1}), we observe that for any value of $\Gamma z$
compound mode $ab_{1}$ is higher order entangled. However, Hillery-Zubairy's
higher order entanglement criteria (\ref{hoe-criteria})-(\ref{hoe-criteria-1})
could not show any signature of higher order entanglement in compound
modes $ab_{2}$ and $b_{1}b_{2}.$ This is not surprising as Hillery-Zubairy's
criteria are only sufficient not necessary and we have already seen
that these criteria fail to detect lower order entanglement present
in compound modes $ab_{2}$ and $b_{1}b_{2}.$

There exists another way to study higher order entanglement. To be
precise, all multi-mode entanglement are essentially higher order
entanglement. As there are 3 modes in the coupler studied here, we
may also investigate the existence of three-mode entanglement. We
have already noted that a 3-mode pure state that violates (\ref{eq:fully ent2})
(i.e., satisfies $\langle N_{a}\rangle\langle N_{b_{1}}\rangle\langle N_{b_{2}}\rangle-|\langle ab_{1}b_{2}\rangle|^{2}<0)$
and simultaneously satisfies either (\ref{eq:fully enta 0}) or (\ref{eq:fully enta 1})
is a fully entangled state. Now using (\ref{eq:2})-(\ref{eq:3})
and (\ref{eq:tripartite ent1})-(\ref{eq:fully ent2}) we obtain following
relations for $m=n=l=1$: 

\begin{equation}
E_{a|b_{1}b_{2}}^{1,1,1}=-E_{a|b_{1}b_{2}}^{'1,1,1}=E_{ab_{2}|b_{1}}^{1,1,1}=-E_{ab_{2}|b_{1}}^{'1,1,1}=|\gamma|^{2}E_{ab_{1}}^{1,1},\label{eq:relation1}\end{equation}
 \begin{equation}
E_{ab_{1}|b_{2}}^{1,1,1}=E_{ab_{1}|b_{2}}^{'1,1,1}=0,\label{eq:relation2}\end{equation}
and \begin{equation}
\langle N_{a}\rangle\langle N_{b_{1}}\rangle\langle N_{b_{2}}\rangle-|\langle ab_{1}b_{2}\rangle|^{2}=-|\gamma|^{2}E_{ab_{1}}^{1,1}.\label{eq:relation3}\end{equation}
From (\ref{eq:relation1}) we can see that three modes of the coupler
is not bi-separable in the form $a|b_{1}b_{2}$ and $ab_{2}|b_{1}$
for any value of $\Gamma z>0.$ Further, Eqn. (\ref{eq:relation3})
and positive regions of $E_{ab_{1}}^{1,1}$ shown in Fig. \ref{fig:ent-hz1-II}
show that the three modes of the coupler are not fully separable.
However, present solution does not show signature of fully entangled
3-mode state as (\ref{eq:relation2}) does not show entanglement between
coupled mode $ab_{1}$ and mode $b_{2}.$ To be specific, we observed
3-mode (higher order) entanglement, but could not observe signature
of fully entangled 3-mode state. However, here we cannot conclude
whether the three modes of the coupler are fully entangled or not
as the criteria used here are only sufficient. 

We have already observed different signatures of nonclassicality in
asymmetric nonlinear optical coupler of our interest. If we now closely
look into all the analytic expressions of signatures of nonclassicality
provided here through Eqs. (\ref{eq:asq-1})-(\ref{eq:relation3})
we can find an interesting symmetry: all the non-vanishing expressions
of signatures of nonclassicality are proportional to $|\gamma|.$
Thus we may conclude that within the domain of validity of the present
solution, in the spontaneous process we would not observe any of the
nonclassical characters that are observed here in stimulated case.

\section{{\normalsize Conclusions \label{sec:Conclusions}}}

We have observed various types of higher order nonclassicality in
fields propagating through a codirectional asymmetric nonlinear optical
coupler prepared by combining a linear wave guide and a nonlinear
(quadratic) wave guide operated by second harmonic generation. The
observations are elaborated in Section \ref{sec:Nonclassicality-in-codirectional}.
In brief, we have observed higher order (amplitude squared) squeezing,
higher order antibunching and higher order entanglement. None of these
higher order nonclassical phenomena were reported in earlier studies
on the codirectional asymmetric nonlinear optical coupler (\cite{Mandal-Perina}
and references therein). In fact, till date neither entanglement nor
higher order nonclassicalities are systematically studied in optical
couplers other than the Kerr coupler. The method followed in the present
paper is quite general and it can be extended easily to the other
type of couplers, such as contradirectional asymmetric nonlinear coupler,
codirectional and contradirectional Raman and Brillouin coupler \cite{Raman and Brilloin}
and parametric coupler \cite{parametric}. Further, it is even possible
to investigate the existence of Hong-Mandel \cite{Hong-Mandel1,HOng-mandel2}
type higher order squeezing and Agarwal-Tara parameter $A_{n}$ \cite{Agarwal-Tara}
for higher order nonclassicality using the present approach. It is
also possible to study lower order and higher order steering using
the present approach and the strategy adopted in Ref. \cite{He-steering}.
However, we have not investigated steering as recently it is shown
that every pure entangled state is maximally steerable \cite{steering in pure state}.
Since the combined states of three modes of asymmetric codirectional
optical coupler is a pure state, the findings of Ref. \cite{steering in pure state}
and the intermodal entanglement observed in the present paper implies
that the compound modes $ab_{1}$ is maximally steerable. The importance
of entanglement and steering in various applications of quantum computing
and quantum communication and the easily implementable structure of
the coupler studied here indicate the possibility that the entangled
states generated through the coupler of the present form would be
useful in various practical purposes. 

\textbf{Acknowledgment:} K. T. and A. P. thank the Department of Science
and Technology (DST), India, for support provided through DST project
No. SR/S2/LOP-0012/2010. A. P. also thanks Operational Program Education
for Competitiveness-European Social Fund project CZ.1.07/2.3.00/20.0017
of the Ministry of Education, Youth and Sports of the Czech Republic.
A. P. and J. P. thank the Operational Program Research and Development
for Innovations - European Regional Development Fund project CZ.1.05/2.1.00/03.0058
of the Ministry of Education, Youth and Sports of the Czech Republic.
Further, A. P. thanks J. ${\rm \check{Perina}}$ Jr. for some helpful
technical discussions.


\begin{thebibliography}{56}
\bibitem{CV-qkd-hillery}M. Hillery, Phys. Rev. A \textbf{61}, 022309
(2000).

\bibitem{teleportation of coherent state}A. Furusawa, J. L. Sorensen,
S. L. Braunstein, C. A. Fuchs, H. J. Kimble and E. S. Polzik, Science
\textbf{282}, 706 (1998).

\bibitem{antibunching-sps}Z. Yuan, B. E. Kardynal, R. M. Stevenson,
A. J. Shields, C. J. Lobo, K. Cooper, N. S. Beattie, D. A. Ritchie
and M. Pepper, Science \textbf{295}, 102 (2002).

\bibitem{Ekert protocol}A. Ekert, Phys. Rev. Lett. \textbf{67}, 661
(1991).

\bibitem{Bennet1993}C. H. Bennett, G. Brassard, C. Crepeau, R. Jozsa,
A. Peres and W. K. Wootters, Phys. Rev. Lett. \textbf{70}, 1895 (1993).

\bibitem{densecoding}C. H. Bennett and S. J. Wiesner, Phys. Rev.
Lett. \textbf{69}, 2881 (1992).

\bibitem{pathak-perina}A. Pathak, J. K\u{r}epelka and Jan Pe\v{r}ina,
Phys. Lett. A \textbf{377}, 2692 (2013).

\bibitem{pathak-PRA}B. Sen, S. K. Giri, S. Mandal, C. H. R. Ooi and
A. Pathak, Phys. Rev. A \textbf{87}, 022325 (2013).

\bibitem{Nature09} J. C. F. Matthews, A. Politi, A. Stefanov and
J. L. O\textquoteright{}Brien, Nature Photonics \textbf{3}, 346 (2009).

\bibitem{Mandal-Midda} P. Mandal and S. Midda, Optik \textbf{122},
1795 (2011).

\bibitem{perina-review} J. Perina Jr. and J. Perina, {}``Quantum
statistics of nonlinear optical couplers '' in: E. Wolf (Ed.), Progress
in Optics, \textbf{41}, Elsevier, Amsterdam, 361 (2000). 

\bibitem{Kerr-1}F. A. A. El-Orany, M. S. Abdalla and J. Perina, Eur.
Phys. J. D, \textbf{33}, 453 (2005).

\bibitem{kerr-2}N. Korolkova, J. Perina, Opt. Commun. \textbf{136},
135 (1997). 

\bibitem{Kerr-3}J. Fiurasek, J. Krepelka, J. Perina, Opt. Commun.
\textbf{167}, 115 (1999). 

\bibitem{kerr-4}G. Ariunbold, J. Perina, Opt. Commun. \textbf{176},
149 (2000).

\bibitem{kerr-5}N. Korolkova, J. Perina, J. Mod. Opt.\textbf{ 44},
1525 (1997). 

\bibitem{Raman and Brilloin} J. Perina Jr. and J. Perina, Quantum
Semiclass. Opt. \textbf{9}, 443 (1997).

\bibitem{parametric}N. Korolkova, J. Perina, Opt. Commun. \textbf{137},
263 (1997). 

\bibitem{assymetric-strong-pump}J. Perina and J. Perina Jr., Quantum
Semiclass. Opt. \textbf{7}, 541 (1995).

\bibitem{assymetric-2}J. Perina, J. Mod. Opt. \textbf{42}, 1517 (1995). 

\bibitem{Mandal-Perina}S. Mandal and J. Perina, Phys. Lett. A \textbf{328,}
144 (2004).

\bibitem{co-and-contra}J. Perina, J. Perina Jr., J. Mod. Opt. \textbf{43},
1951 (1996). 

\bibitem{contra-1} J. Perina and J. Perina Jr., Quantum Semiclass.
Opt. \textbf{7}, 849 (1995).

\bibitem{nonclassical-input1}J. Perina, J. Bajer, J. Mod. Opt. \textbf{42},
2337 (1995). 

\bibitem{nonclasssical input2}L. Mista Jr. and J. Perina, Czechoslovak
Journal of Physics, \textbf{47}, 629 (1997) .

\bibitem{co and contra with phase mismatch} J. Perina and J. Perina
Jr., Quantum Semiclass. Opt. \textbf{7}, 863 (1995).

\bibitem{leonoski1} A. Kowalewska-Kud\l{}aszyk and W. Leonski, J.
Opt. Soc. B \textbf{26}, 1289 (2009) .

\bibitem{thermal ent-kerr} M. R. Abbasi and M. M. Golshan, Physica
A \textbf{392}, 6161 (2013).

\bibitem{kerr-lionoski-miranowicz}W. Leonski and A. Miranowicz, J.
Opt. B: Quantum Semiclass. Opt. \textbf{6}, S37 (2004).

\bibitem{GBS} A. Kowalewska-Kud\l{}aszyk, W. Leonski and J. Perina
Jr., Phys. Scr. \textbf{T147}, 014016 (2012).

\bibitem{higher-order} F. A. A. El-Orany and J. Perina, Phys. Lett.
A \textbf{333}, 204 (2004).

\bibitem{bsen1} B. Sen and S. Mandal, J. Mod. Opt. \textbf{52}, 1789
(2005).

\bibitem{generalized-higher order}A. Verma and A. Pathak, Phys. Lett.
A \textbf{374}, 1009 (2010).

\bibitem{Maria-PRA-1}A. Allevi, S. Olivares and M. Bondani, Phys.
Rev. A \textbf{85}, 063835 (2012).

\bibitem{Maria-2}A. Allevi, S. Olivares and M. Bondani, Int. J. Quant.
Info. \textbf{8}, 1241003 (2012).

\bibitem{higher-order-PRL}M. Avenhaus, K. Laiho, M. V. Chekhova and
C. Silberhorn, Phys. Rev. Lett \textbf{104}, 063602 (2010).

\bibitem{HOAwithMartin}A. Pathak and M. Garcia, Applied Physics B
\textbf{84}, 484 (2006).

\bibitem{HIlery-amp-sq}M. Hillery, Phys. Rev. A \textbf{36}, 3796
(1987).

\bibitem{Hong-Mandel1}C. K. Hong, L. Mandel, Phys. Rev. Lett. \textbf{54,}
323 (1985). 

\bibitem{HOng-mandel2}C. K. Hong, L. Mandel, Phys. Rev. A \textbf{32},
974 (1985). 

\bibitem{Perina-Book}J. Perina, Quantum Statistics of Linear and
Nonlinear Optical Phenomena, Kluwer, Dordrecht (1991).

\bibitem{C T Lee}C. T. Lee, Phys. Rev. A \textbf{41}, 1721 (1990).

\bibitem{HOAis not rare}P. Gupta, P. Pandey and A. Pathak, J. Phys.
B \textbf{39}, 1137 (2006).

\bibitem{with bishu Arxive}S. K. Giri, B. Sen, C. H. R. Ooi, and
A. Pathak, Phys. Rev. A \textbf{89}, 033628 (2014). 

\bibitem{NJP-GS-Ashoka} G. S. Agarwal and A. Biswas, New J. Phys.
\textbf{7}, 211 (2005).

\bibitem{duan} L. M. Duan, G. Giedke, J. I. Cirac and P. Zoller,
Phys. Rev. Lett. \textbf{84}, 2722 (2000).

\bibitem{HZ-PRL} M. Hillery and M. S. Zubairy, Phys. Rev. Lett. \textbf{96},
050503 (2006).

\bibitem{HZ2007} M. Hillery and M. S. Zubairy, Phys. Rev. A \textbf{74},
032333 (2006).

\bibitem{HZ2010} M. Hillery, H. T. Dung and H. Zheng, Phys. Rev.
A \textbf{81}, 062322 (2010).

\bibitem{higher order multiparty1} A. Zeilinger, M. A. Horne and
D. M. Greenberger, NASA Conf. Publ. \textbf{3135}, 51 (1992).

\bibitem{higher order multyparty2}J.-W. Pan, M. Daniell, S. Gasparoni,
G. Weihs and A. Zeilinger, Phys. Rev. Lett. \textbf{86}, 4435 (2001).

\bibitem{higher order multyparty3}A. Mair, A. Vaziri, G. Weihs and
A. Zeilinger, Nature \textbf{412}, 313 (2001).

\bibitem{Ent condition-multimode}Z.-G. Li, S.-M. Fei, Z.-X. Wang
and K. Wu, Phys. Rev. A \textbf{75}, 012311 (2007).

\bibitem{Agarwal-Tara}G. S. Agarwal and K. Tara, Phys. Rev. A \textbf{46},
485(1992).

\bibitem{He-steering}Q. Y. He, P. D. Drummond, M. K. Olsen and M.
D. Reid, Phys. Rev. A \textbf{86}, 023626 (2012).

\bibitem{steering in pure state}P. Skrzypczyk, M. Navascues and D.
Cavalcanti, arXiv:1311.4590 (2013).
\end{thebibliography}
\end{document}